\documentclass[twocolumn]{aastex631}

\usepackage{longtable}
\usepackage{CJKutf8, color}

\renewcommand{\figurename}{Fig.}

\received{---}
\revised{---}
\accepted{---}

\newcommand{\kanji}[1]{\begin{CJK}{UTF8}{ipxm}(#1)\end{CJK}}

\shorttitle{Impact of Neutrino Cooling on Type-I X-ray Bursts and X-ray Superbursts}
\shortauthors{Dohi et al.}

\begin{document}

\title{Impacts of the direct URCA and Superfluidity inside a Neutron Star on Type-I X-Ray Bursts and X-Ray Superbursts}

\correspondingauthor{Akira Dohi}
\email{dohi@phys.kyushu-u.ac.jp}

\author[0000-0001-8726-5762]{A.~Dohi \kanji{土肥明}}
\affiliation{Department of Physics, Kyushu University, Fukuoka 819-0395, Japan}
\affiliation{Interdisciplinary Theoretical and Mathematical Sciences Program (iTHEMS), RIKEN, Wako, Saitama 351-0198, Japan}

\author[0000-0002-0842-7856]{N.~Nishimura \kanji{西村信哉}}
\affiliation{Astrophysical Big Bang Laboratory, Cluster for Pioneering Research, RIKEN, Wako, Saitama 351-0198, Japan}
\affiliation{Nishina Center for Accelerator-Based Science, Wako, Saitama 351-0198, Japan}
\affiliation{Division of Science, National Astronomical Observatory of Japan, Mitaka 181-8588, Japan}

\author[0000-0002-3239-2921]{H.~Sotani \kanji{祖谷元}}
\affiliation{Astrophysical Big Bang Laboratory, Cluster for Pioneering Research, RIKEN, Wako, Saitama 351-0198, Japan}
\affiliation{Interdisciplinary Theoretical and Mathematical Sciences Program (iTHEMS), RIKEN, Wako, Saitama 351-0198, Japan}

\author[0000-0003-0943-3809]{T.~Noda \kanji{野田常雄}}
\affiliation{Department of Education and Creation Engineering, Kurume Institute of Technology, Kurume, Fukuoka 830-0052, Japan}

\author{He-Lei Liu}
\affiliation{School of Physical Science and Technology, Xinjiang University, Urumqi 830046, China}

\author[0000-0002-7025-284X]{S.~Nagataki \kanji{長瀧重博}}
\affiliation{Astrophysical Big Bang Laboratory, Cluster for Pioneering Research, RIKEN, Wako, Saitama 351-0198, Japan}
\affiliation{Interdisciplinary Theoretical and Mathematical Sciences Program (iTHEMS), RIKEN, Wako, Saitama 351-0198, Japan}

\author{M.~Hashimoto \kanji{橋本正章}}
\affiliation{Department of Physics, Kyushu University, Fukuoka 819-0395, Japan}



\begin{abstract}
We investigate the impacts of neutrino cooling mechanism inside the neutron star (NS) core on the light curves of type-I X-ray bursts and X-ray superbursts. From several observations of NS thermal evolution, physical processes of fast neutrino cooling, such as the direct Urca (DU) process, are indicated. They significantly decrease the surface temperature of NSs, though the cooling effect could be suppressed by nucleon superfluidity. In the present study, focusing on the DU process and nucleon superfluidity, we investigate the effects of NS cooling on the X-ray bursts using a general-relativistic stellar-evolution code.
We find that the DU process leads find the longer recurrence time and the higher peak luminosity, which could be obstructed by the neutrons superfluidity. We also apply our burst models to the comparison with {\it Clocked burster} GS 1826$-$24, and to the recurrence time of superburst triggered by carbon ignition. These effects are significant within a certain range of binary parameters and uncertainty of the NS equation of state.
\end{abstract}



\section{Introduction}
\label{sec:intro}

Thermonuclear explosions triggered by hydrogen or helium (H/He) burning at the mass accreted layer on the neutron star (NS) have been observed as type-I X-ray bursts. The type-I X-ray burst is initiated from the triple-$\alpha$ reaction, occurring around $T=0.2~{\rm GK}$. At low accretion rates of $\dot{M}\lesssim10^{-9}~M_{\odot}~{\rm yr^{-1}}$~\citep{1981ApJ...247..267F}, the most of hydrogen is consumed and transformed to helium. Thus, pure-helium burning occurs and results in the rapid increase to near the Eddington luminosity. Since there are few protons in this case, the next reaction after 3$\alpha$ reaction is not proton capture but $\alpha$ capture of carbons. Due to a series of ($\alpha,\gamma$) reactions, even-even nuclei up to ${}^{56}{\rm Ni}$ are synthesized and finally the burst phase is terminated.

At moderate accretion rates of $\dot{M}\gtrsim10^{-9}~M_{\odot}~{\rm yr^{-1}}$ which we consider in this paper, on the other hand, the hot CNO cycle occurs after 3$\alpha$ reaction. The rapid temperature increase by the nuclear burning results in the linear increase of the luminosity. When the temperature exceeds $0.5~{\rm GK}$, the breakthrough of the hot CNO cycle occurs and the luminosity reaches the maximum, typically $10^{38}~{\rm erg~s^{-1}}$. Additionally, nucleosynthesis by the $\alpha p$ and $rp$ processes begin, where proton-rich heavy nuclei up to $A \sim 106$~\citep{2001PhRvL..86.3471S} are synthesized. Then, the luminosity decreases exponentially with $t \sim 100~{\rm s}$ and finally the burst phase is terminated.

The physics of X-ray bursts is complicated as they involve in not only the nuclear burning but also high dense matter of NSs and X-ray binary parameters. Modeling of observed burst light curves, therefore, is subject to probe the underlying properties of NS equations of state (EOSs), the accretion rate, and the composition of accreting matter. In addition, the uncertainty of nuclear reaction rates especially for $\alpha p$ and $rp$ processes affects the modeling of burst light curves \citep{2021PhRvL.127q2701H, 2021arXiv211013676L, 2021arXiv210711552L, 2021arXiv211014558M}. The dependence of the above physical parameters on X-ray burst light curves has been investigated by many theoretical works \citep{2004ApJS..151...75W, 2007ApJ...671L.141H, 2016ApJ...830...55C, 2018ApJ...860..147M, 2019ApJ...872...84M, 2020MNRAS.494.4576J, 2021arXiv210711552L}. 

Currently, 115 X-ray bursters have been observed \citep[see, e.g.,][]{2020ApJS..249...32G}. Although there are various light-curve profiles, some bursters show the almost same pattern. They are called {\it Clocked bursters} and can provide a standard case to constrain theoretical model. In particular, bursts from GS 1826$-$24 are the most commonly used as such references. Based on numerical models covering a wide range of burst parameters, \cite{2020MNRAS.494.4576J} constrain burst models by the observations of GS 1826$-$24, e.g., $Z_{\rm CNO}=0.01^{+0.005}_{-0.004}$, and $\dot{M}_{-9}=1.5$--$3.0$, where $Z_{\rm CNO}$ is the initial metallicity and $\dot{M}_{-9}$ is the accretion rate normalized by $10^{-9}~M_{\odot}~{\rm yr^{-1}}$.

Previous theoretical modelings of X-ray bursters, however, have mostly focused on the accreted layers, ignoring detailed physics inside the NS. For such models, the effects of the NS cooling and heating are imposed on $Q_b$, the energy change though the crust surface, as the boundary condition. Although we can easily investigate parameter dependence of the heating strength with $Q_b$, realistic physics in the NS core are not fully considered. To overcome this, we have employed the original general relativistic (GR) evolutionary code \citep[see,][]{1984ApJ...278..813F, 2020PTEP.2020c3E02D}. Based on X-ray burst models with realistic EOSs, we have shown that the neutrino emission from the NS significantly affects the light curves \citep{2021ApJ...923...64D}: the recurrence time ($\Delta t$) and peak luminosity ($L_{\rm peak}$).

In our previous study \citep{2021ApJ...923...64D}, in order to focus on the EOS and NS mass dependence on the light curve, we took into account only the slow cooling scenario, i.e., the effects of the fast cooling and the nucleon superfluidity are ignored. However, recent NS thermal evolution observations in particular Cassiopeia~A \citep{2011PhRvL.106h1101P, 2011MNRAS.412L.108S}, indicate {\it minimal cooling scenario} \citep{2004ApJS..155..623P, 2004A&A...423.1063G}, where we must consider the combination of slow neutrino cooling processes and the pair breaking and formation (PBF) process due to nucleon superfluid state. In addition, some cold NSs observations require even faster cooling by e.g. the nucleon Direct Urca (DU) process~(e.g., see Section 1 in \cite{2022IJMPE..3150006D}).

On the other hand, the cooling effects on the burst behavior have been also widely discussed in the context of superbursts, which is longer  duration burst (1000 s) than mixed H/He burst (10--100 s)~\citep{2003A&A...411L.487I,2017nuco.confb0304S,2021PASJ...73.1405I}. Superburst is thought to be a thermonuclear runaway initially caused by carbon unstable burning~\citep{2006ApJ...646..429C}. Generally, since the depth of the ignition of carbons is deeper than that of the H/He ignition, superburst should be affected by the neutrino losses at the relatively deeper layer inside NSs compared with usual X-ray bursts~\citep{2004ApJ...614L..57B,2006ApJ...646..429C,2007ApJ...662.1188G,2016ApJ...831...13D}. In particular, the impacts of neutrino cooling on the recurrence time of superburst strongly depend on the thermal neutrino losses (e.g., \cite{2007ApJ...662.1188G} for nucleon-pair bremsstrahlung and crust PBF process). Thus, the DU process may also affect the depth of carbon ignition and the observed recurrence time of superburst, although no previous studies examine the effect of the DU process. Not only the neutrino cooling effects but also the reaction rate uncertainties of carbon burning on the superburst behavior have been discussed~(e.g, \cite{2009ApJ...702..660C}).

In this study, we investigate the impacts of neutrino cooling on X-ray bursts and superburst within our multi-zone models with the entire NSs considered~\citep{2020PTEP.2020c3E02D}. In particular, we discuss the impacts of the DU process and nucleon superfluidity on both the mixed H/He bursts and on the carbon ignition.

This paper is organized as follows. In Section~\ref{sec:setup}, we describe our numerical methods with an emphasis on the neutrino cooling processes and the initial models for X-ray burst calculations. In Section~\ref{sec:result}, we show the effect of neutrino cooling on the light curves of Type-I X-ray bursts, focusing on GS 1826$-$24. In section~\ref{sec:c12c12}, we discuss the impacts of NS cooling on the carbon ignition, which results in the X-ray superburst, considering uncertainty of the reaction rate. We finally give a concluding remarks in Section~\ref{sec:conc}.

\begin{figure}[t]
    \centering
    \includegraphics[width=\linewidth]{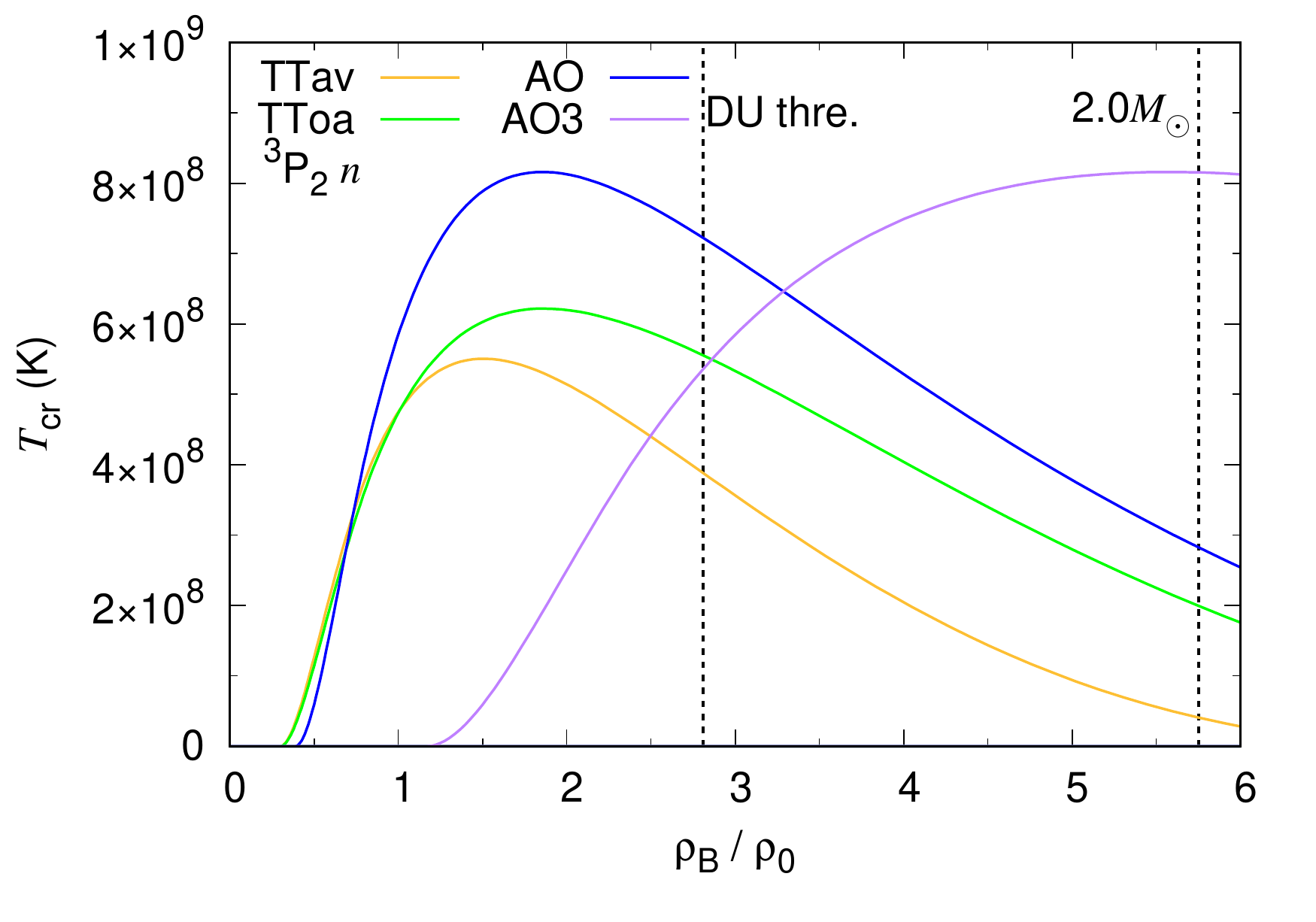}
    \caption{Critical temperature with different neutron ${}^3P_2$ superfluid models is shown as a function of normalized baryon density with $\rho_{0}=2.57\times10^{14}~{\rm g~cm^{-3}}$, which is the nuclear saturation density for LS220 EOS. We also draw dotted lines at the threshold density of the DU process and the central density of $2.0~M_{\odot}$ stars.}
    \label{fig:sf}
\end{figure}

\section{Numerical Setup}
\label{sec:setup}
\subsection{Input Physics}
\label{subsec:inp}

To calculate burst light curves, we use the numerical code developed in our previous work~\citep{2020PTEP.2020c3E02D}. Our formulation is based on one-dimensional GR hydrostatic equilibrium condition with 88-nuclei approximated reaction network for mixed H/He burning. In this study we especially adopt the EOS proposed by \cite{1991NuPhA.535..331L} with the incompressibility of $K=220~{\rm MeV}$ (LS220).

With the LS220 EOS, the DU process works for the NS models with $M_{\rm NS}\gtrsim1.35~M_{\odot}$ (e.g., \cite{2019PTEP.2019k3E01D}), and the expected maximum mass exceeds the $2 M_{\odot}$ observations~\citep{2010Natur.467.1081D,2013Sci...340..448A,2020NatAs...4...72C}. To examine the effect of DU process, we focus only on the NS model with $M_{\rm NS}=2.0~M_{\odot}$ ($R_{\rm NS}=11.3~{\rm km}$). We note that the stellar model considered in this study agrees with the theoretical constraints obtained from {\it best-fit} models of GS 1826$-$24 \citep{2020MNRAS.494.4576J}, i.e., $M_{\rm NS}\gtrsim1.7~M_{\odot}$ and $R_{\rm NS}=11.3\pm1.3~{\rm km}$, (but see also \cite{2012ApJ...749...69Z} as a counterexample). For the heating mechanism inside NSs, the effective process is the crustal heating, which is the energy release due to the non-equilibrium nuclear reactions of accreted matter such as the electron capture, neutron emission, and pycnonuclear reactions. For their reaction rates, we simply adopt the conventional energy generation rates proposed by \cite{2008A&A...480..459H}.

Regarding the neutrino cooling processes, we consider the DU process as a fast cooling process, together with the minimal cooling processes. Once the DU process works, the internal temperature of NS  drastically decreases. On the other hand, when the DU process is forbidden, the dominant cooling processes are modified URCA, bremsstrahlung, and the PBF process of nucleon superfluidity~\citep{2001PhR...354....1Y}. For isolated NSs, the PBF process significantly decreases their surface temperature within $t=10^{1-3}~{\rm yrs}$~\citep{2004ApJS..155..623P}, but this process does not affect the temperature observation of old accreting NSs. The PBF process works only when the temperature of the layer declines through the critical temperature $T_{\rm cr}$. This situation hardly occurs in the accreting NSs, and such an additional cooling process is not so effective~\citep{2019A&A...629A..88P}.

Focusing on the DU process, the neutron ${}^3P_2$  superfluidity is known to decrease the neutrino emissivity in low-temperature regions~\citep{1972PThPh..48.1517T}. Therefore, we adopt three kinds of neutron ${}^3P_2$ superfluid models, which are different density dependence in $T_{\rm cr}$, i.e., TTav, TToa~\citep{2004PThPh.112...37T,2015PhRvC..91a5806H}, AO~\citep{1985NuPhA.442..163A}, and AO3 for which we multiply the density in AO model by a factor of three. That is, the maximum critical temperature in AO3 is the same as in AO, but the ``peak" density at the maximum critical temperature is different from each other. We show their superfluid models in Fig.~\ref{fig:sf}. In all regions above the saturation density, AO and AO3 are the models with the highest critical temperature, while TTav is the model with the lowest critical temperature among four neutron ${}^3P_2$ models considered in this study. Meanwhile, for ${}^1S_0$-state neutrons/protons superfluidity, we simply fix them as Chen-Clark-Dav{\'e}-Khodel (CCDK) model~\citep{1993NuPhA.555...59C,2015PhRvC..91a5806H}, where the maximum critical temperature is $T_{\rm cr}\simeq6\times10^9~{\rm K}$ at $\rho=0.56\rho_{\rm 0}$ for neutrons and $1.1\rho_{\rm 0}$ for protons, respectively, where $\rho_{\rm 0}=2.57\times10^{14}~{\rm g~cm^{-3}}$ is the saturation density for LS220 EOS. It is well known that the ${}^1S_0$-state neutrons superfluidity does not contribute to the surface luminosity at all for $t\gtrsim10^2~{\rm yrs}$~(e.g., \cite{2009ApJ...707.1131P}). For the protons superfluidity, the DU process could be suppressed, whose effect is generally weaker than the neutron ${}^3P_2$ superfluidity as we show later. In Table \ref{tab:correspondence} we summarize the name of models and the effects taken into account in this study.

\begin{table}[th]
    \centering
    \caption{Correspondence between the name of models and effects taken into account.}
    \begin{tabular}{c|ccc}
    \hline\hline
    Model & direct URCA & ${}^1S_0 (n/p)$ & ${}^3P_2 (n)$  \\
    \hline
    {\tt Normal}  & $\checkmark$ & -- & -- \\
    {\tt CCDK}    & $\checkmark$ & $\checkmark$ & -- \\
    {\tt CCDK+TTav} & $\checkmark$ & $\checkmark$ & TTav \\
    {\tt CCDK+TToa} & $\checkmark$ & $\checkmark$ & TToa \\
    {\tt CCDK+AO}   & $\checkmark$ & $\checkmark$ & AO \\
    {\tt CCDK+AO3}   & $\checkmark$ & $\checkmark$ & AO($\rho_{\rm B} \times 3$) \\
    {\tt No DU} & -- & -- & -- \\
    \hline\hline
    \end{tabular}
    \label{tab:correspondence}
\end{table}

\begin{figure}[t]
    \centering
    \includegraphics[width=\linewidth]{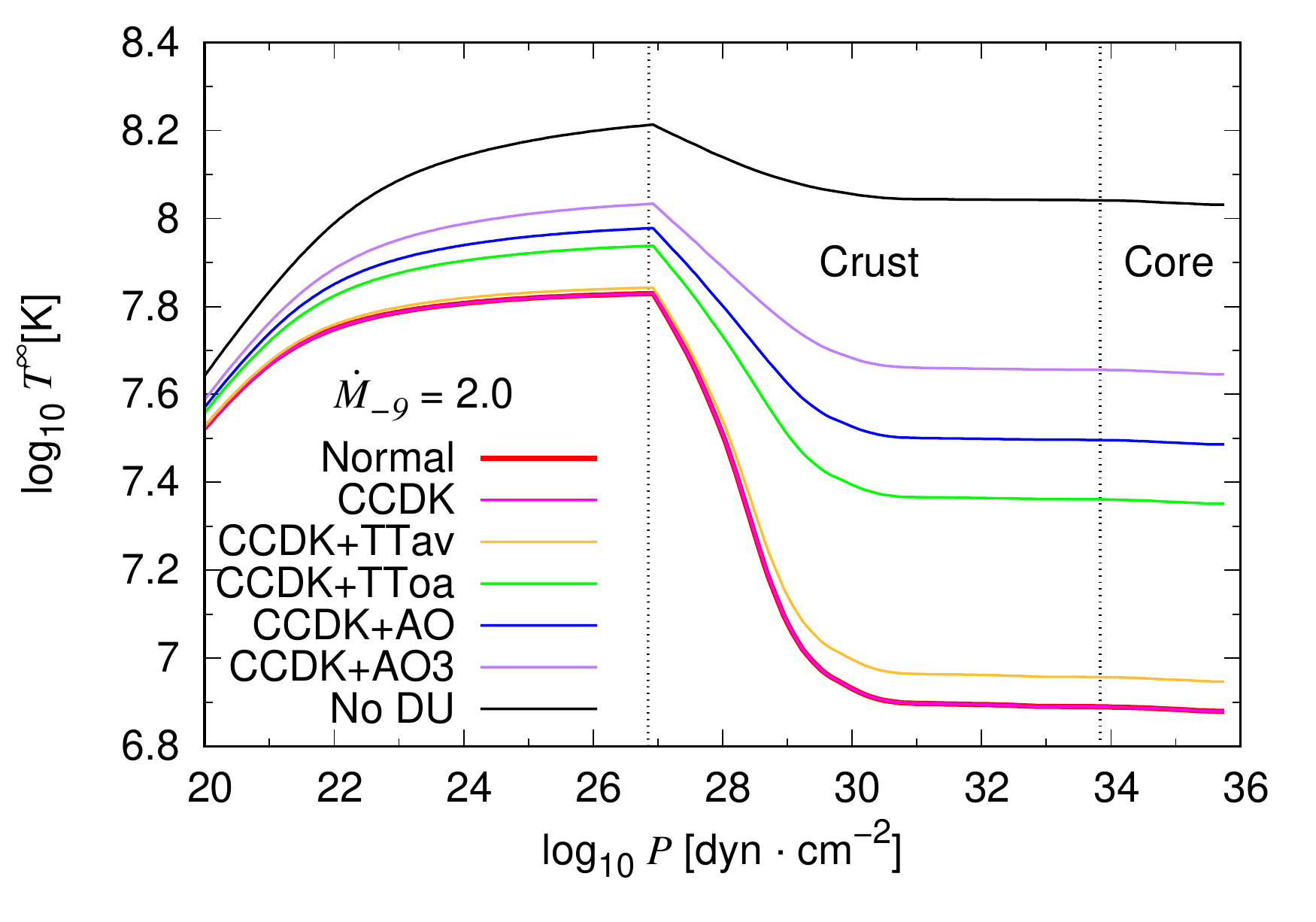}
    \caption{Redshifted temperature structure in steady state with $\dot{M}_{-9}=2.0$. {\tt Normal}(red) indicates the model without any superfluidity and {\tt CCDK}(magenta) without only neutron ${}^3P_2$ superfluidity. We note that the result with {\tt Normal} almost overlaps with that with {\tt CCDK}.  Other models with all kinds of superfluidity are as follows: TTav(yellow), TToa(green), and AO(blue) for neutron ${}^3P_2$ superfluid models, respectively. We draw two vertical lines at the envelop-crust and crust-core boundaries, whose densities respectively correspond to $10^{9}~{\rm g~cm^{-3}}$ and  the saturation density $\rho_{0}$. In addition, we also show the result without the direct URCA labeled by {\tt No DU} for reference. }
    \label{fig:steady}
\end{figure}

\subsection{Initial Thermal Profiles of Accreting NSs}
\label{subsec:init}

To prepare an initial thermal profiles, we calculate thermal evolution of accreting NSs without nuclear burning, adopting the same procedure as in \cite{2021PhRvD.103f3009L}, but we also consider the homogeneous compressional heating, not only the non-homogeneous one as explained by \cite{2018IJMPE..2750067M}. The resultant temperature structure in steady state is considered as the initial thermal profile for calculating the burst light curves. We show such a temperature structure in steady state for various models in Fig.~\ref{fig:steady}, where we assume that $\dot{M}_{-9}=2.0$. Comparing to Fig.~\ref{fig:sf}, one can recognize that the steady-state NS maintains higher temperature, as the critical temperature or the peak density are higher for superfluid models. That is, the neutron ${}^3P_2$ model with higher critical temperature corresponds to relatively lower neutrino emissivity. This situation is understood as follows, i.e., once the temperature decreases and reaches the critical temperature, the DU process, which is fast cooling process, is effectively blocked by the pair production due to neutrons superfluidity, while the PBF process as an effect of superfulidity is less effective than the DU process.  

\begin{figure*}[t]
    \centering
    \includegraphics[width=\linewidth]{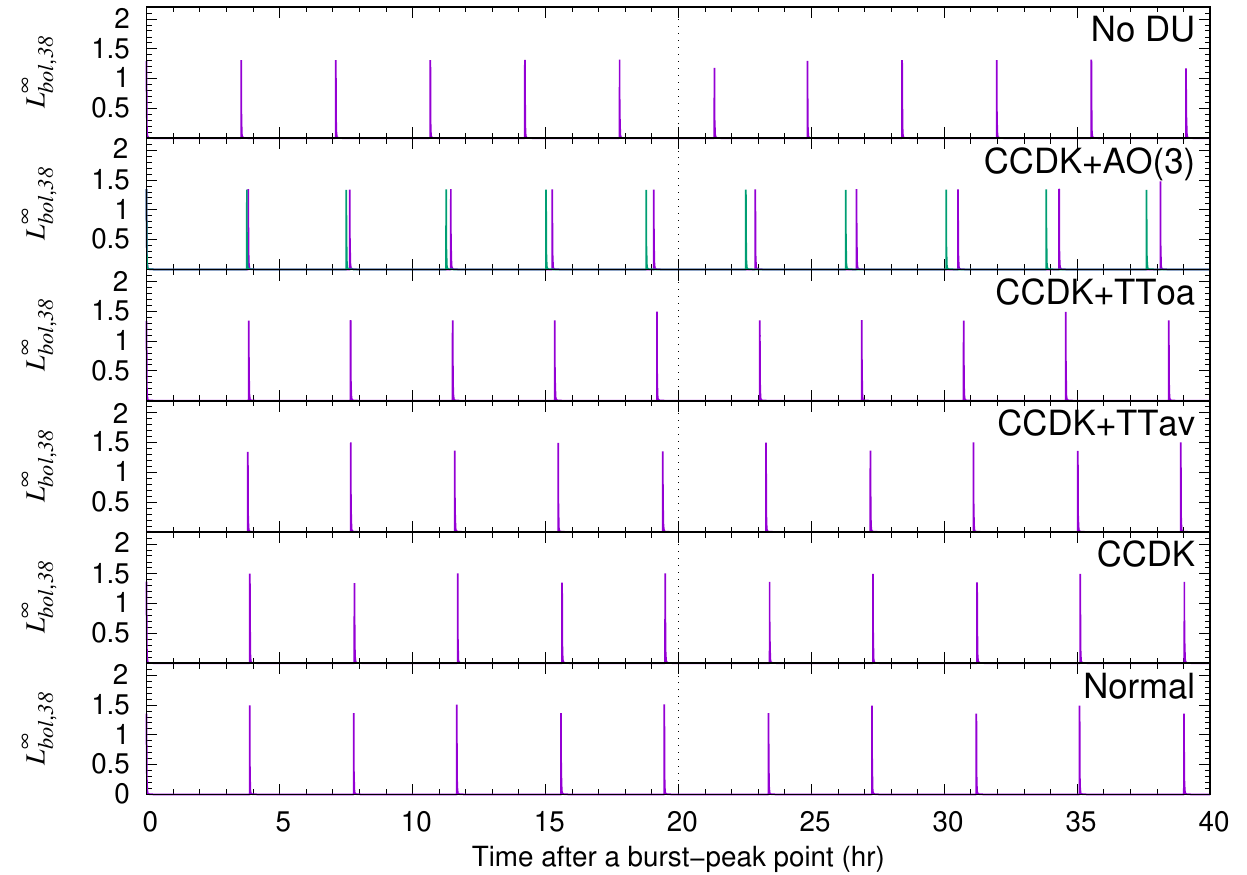}
    \caption{The bolometric luminosity in units of $10^{38}$ erg s$^{-1}$ for the burst sequence occurred during 40 hr for various superfluid models. The meaning of symbol is the same as in  Fig.~\ref{fig:steady}. For AO(purple) and AO3(green), we show their light curves in the same panel. Here, we adopt $\dot{M}_{-9}=2.0$.}
    \label{fig:longlc}
\end{figure*}

\begin{figure*}[t]
    \centering
    \begin{minipage}{0.49\linewidth}
    \includegraphics[width=\linewidth]{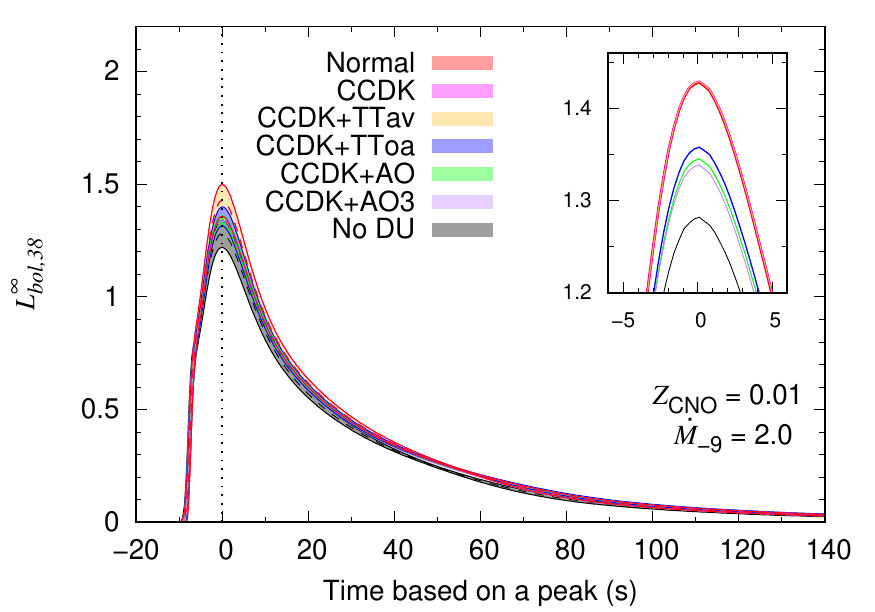}
    \end{minipage}
    \begin{minipage}{0.49\linewidth}
    \includegraphics[width=\linewidth]{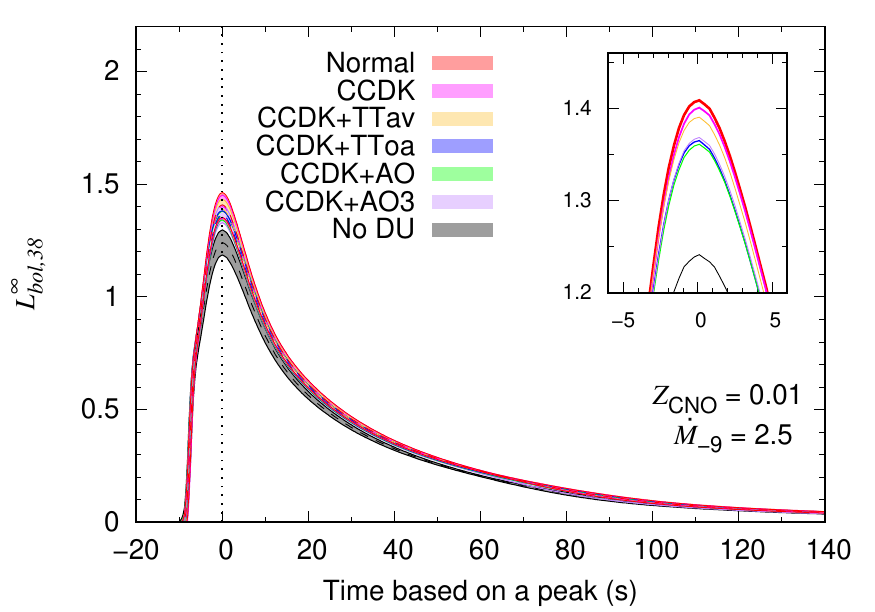}
    \end{minipage}
    \caption{The averaged light curves of the burst phase within $1\sigma$ regions. In the insets, only the most probable light curves are shown for the sake of clarity. We adopt $\dot{M}_{-9}=2.0$ (left) and $2.5$ (right).
    Dashed curves denote the averaged light curves while solid bands include the 1$\sigma$ errors in a numerous bursts.}
    \label{fig:shortlc}
\end{figure*}

\section{Results for type-I X-ray bursts}
\label{sec:result}

In this section, first we discuss the overall of the model dependence of the X-ray burst light curves. We also discuss the relation between the recurrence time and ignition pressure. Then, we compare our results to the concrete observational data in GS 1826$-$24, where we show the dependence of the accretion rate.

\subsection{Model dependence of the light curves}
\label{sec:3a}

We calculate X-ray burst light curves by turning on the nuclear burning which is triggered by mixed H/He burning, adopting the steady-state temperature profile as the initial model for X-ray burst calculation. To eliminate the initial model dependence of burst behavior, i.e., {\it compositional inertia}~\citep{2004ApJS..151...75W}, we discard several dozens of burst profiles within $t\lesssim2\times10^5~{\rm s}$ for all burst models. Then, for the analysis of burst profiles, we select at least 15 successive burst profiles.

In Fig.~\ref{fig:longlc}, we show the superfluid model dependence in long-term light curves, where the first bursts in the selected successive burst profiles for various models are aligned as $t=0$. One observes that the interval between bursts, i.e., the recurrence time $\Delta t$, becomes shorter, as the critical temperature or the peak density in neutron ${^3P_2}$ superfluidity becomes higher. We note that this result is significantly different from the models without DU process ({\tt No DU}). We also find that  the behavior of {\tt CCDK+TTav}, {\tt CCDK} and {\tt Normal} are similar. The recurrence time of other models with strong neutron ${}^3P_2$ superfluidity (TToa, AO, and AO3) are a little shorter (see Table \ref{tab:table}).

\begin{table}[th]
    \centering
    \caption{Averaged recurrence time and ignition pressure with $\dot{M}_{-9}=2.0$ for various models.}
    \begin{tabular}{c|cc}
    \hline\hline
    Model & $\Delta t$ (hr)& $P_{\rm ign}$ (${\rm dyn~cm}^{-2}$)\\
    \hline
      {\tt Normal}  & 3.90  & 4.60$\times10^{22}$ \\
    {\tt CCDK} & 3.90 &  4.60$\times10^{22}$  \\
    {\tt CCDK+TTav} & 3.89 & 4.59$\times10^{22}$\\
    {\tt CCDK+TToa} & 3.83 & 4.53$\times10^{22}$\\
    {\tt CCDK+AO} & 3.81 & 4.50$\times10^{22}$ \\
    {\tt CCDK+AO3} & 3.76 &  4.43$\times10^{22}$\\
    {\tt No DU} & 3.54 & 4.18$\times10^{22}$ \\
     \hline\hline
    \end{tabular}
    \label{tab:table}
\end{table}

Next, to see the model dependence of the peak luminosity, the averaged burst light curves are shown in Fig.~\ref{fig:shortlc}, where the left and right panels correspond to the results with $\dot{M}_{-9}=2.0$ and 2.5, respectively. From this figure, we can classify the models considered in this study into three families, i.e., (i) the lower luminosity ({\tt No DU}), (ii) the higher luminosity ({\tt Normal}, {\tt CCDK}, and {\tt CCDK+TTav}), and (iii) the moderate luminosity ({\tt CCDK+TToa}, {\tt CCDK+AO} and {\tt CCDK+AO3}). Since {\tt Normal} is quite similar to {\tt CCDK} with respect to the recurrence time and peak luminosity, we see that the ${}^1S_0$ superfluidity makes little impact on the burst behavior. We additionally see that {\tt CCDK+TTav} (TTav is a weak neutron ${}^3P_2$ superfluid model) is also similar to (but a little smaller than) {\tt Normal} in the burst peak luminosity, while the peak luminosity with {\tt CCDK+TToa} is similar to that with {\tt CCDK+AO} model and those are lower than that with {\tt Normal} model. Furthermore, if the appearance density of superfluidity is higher, the recurrence time and peak luminosity become a little shorter as we compare with AO and AO3. Thus, we find that the neutron ${}^3P_2$ superfluidity lowers the recurrence time and peak luminosity through the suppression of the DU process, although the range of reduction strongly depends on the neutron ${}^3P_2$ superfluidity model.

\subsection{Empirical relation between $\Delta t$ and the ignition pressure}
\label{sec:3b}

\begin{figure}[h]
\centering
 \includegraphics[width=\linewidth]{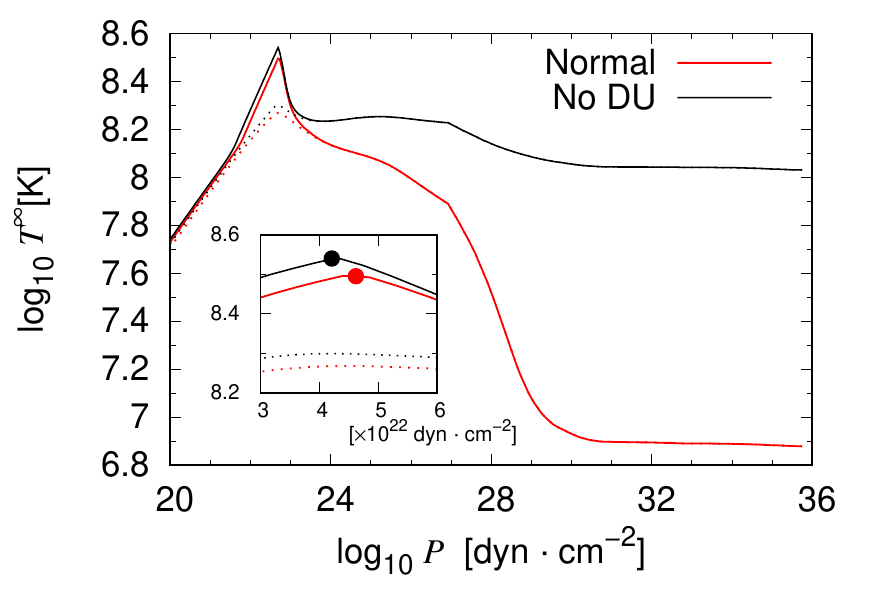}
 \caption{Redshifted temperature structure without and with the DU process. The dotted curves denote those just before the H/He ignition and the solid curves denote just after the H/He ignition, respectively, where we adopt $\dot{M}_{-9}=2.0$.}
 \label{fig:istrc}
\end{figure}

For considering the burst behavior, it is important that  the neutrino cooling process inside the NS decreases the temperature near the typical ignition pressure $P_{\rm ign}\simeq10^{22-23}~{\rm dyn~cm^{-2}}$~\citep{2021ApJ...923...64D} \footnote{We note that the typical pressure for the H/He ignition corresponds to the position, where the temperature becomes $\simeq 0.2$ GK (log$_{10} (T/{\rm [K]})\simeq 8.3$).}. This tendency can be seen even in the case with the DU process, as shown in Fig~\ref{fig:istrc}, but the temperature around ignition pressure is changed in order of $\lesssim 0.2$, depending on the superfluid models. As a result, the position where the temperature becomes $\simeq 0.2$ GK slightly shifts to inner part. Due to the DU process, $P_{\rm ign}$ is increased and indirectly enhance the burst behavior (or the peak luminosity), because the column density $\sigma$ is estimated with the surface gravity $g_s$ via $\sigma=P_{\rm ign}/g_s$, according to the {\it shell-flash} model~\citep{1981ApJ...247..267F}. We note that the difference in the burst behavior in our models comes from only the difference in $P_{\rm ign}$, because the NS mass and radius are fixed in this study. 

In Table~\ref{tab:table}, we list the values of ignition pressure at the time just after the ignitions of hydrogen and helium. As we see, the ignition pressure is maximally changed in order of $\lesssim0.04$. According with the {\it shell-flush} model and assuming the constant accretion rate, the column density $\sigma$ is expressed in two ways~(see also \cite{2007ApJ...662.1188G}):
\begin{eqnarray}
\sigma=\frac{\left(10^{-9}~M_{\odot}~{\rm yr}^{-1}\right)\dot{M}_{-9}}{4\pi R_{\rm NS}^2}\Delta t=P_{\rm ign}/g_{\rm s}~,
\label{eq:eq1}
\end{eqnarray}
where $g_{\rm s}$ is the surface gravity expressed as
\begin{eqnarray}
g_{\rm s} = \frac{GM_{\rm NS}}{R_{\rm NS}^2}\left(1 - \frac{2GM_{\rm NS}}{R_{\rm NS}c^2} \right)^{-1/2}~.
\label{eq:eq3}
\end{eqnarray}
In this equation, $c$ is the speed of light and $G$ is the gravitational constant. With the NS model considered in this study, $g_{\rm s}=3.01\times10^{14}~{\rm cm~s^{-2}}$. Thus, if the NS mass, radius, and accretion rate are fixed, $\Delta t$ should be proportional to $P_{\rm ign}$. In fact, one can see that $P_{\rm ign}/\Delta t$ does not depend on the models considered in this study, i.e., $P_{\rm ign}/\Delta t=1.2\times 10^{22}$~(dyn~cm$^{-2}$/hr), using the values listed in Table \ref{tab:table}. 
That is, one can discuss the recurrence time of X-ray burst via Eq. (\ref{eq:eq1}) even with the effect of superfulidity.

\begin{figure}[t]
    \centering
    \includegraphics[width=\linewidth]{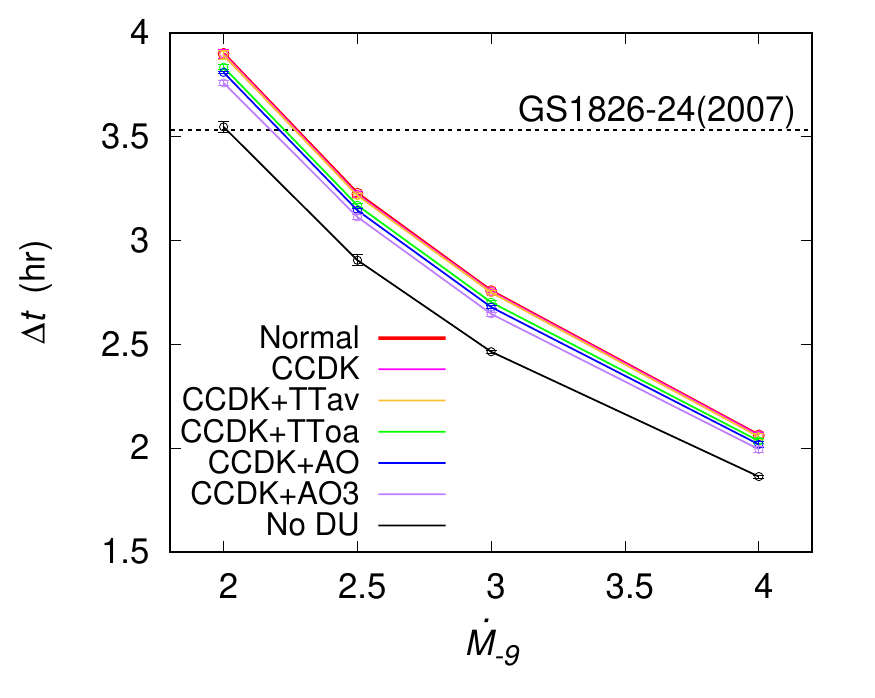}
    \includegraphics[width=\linewidth]{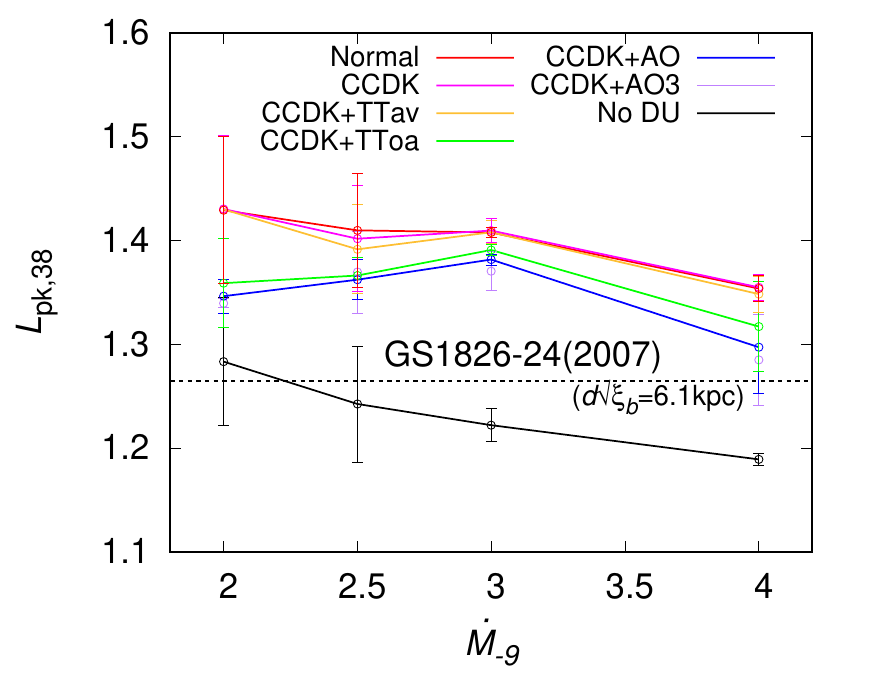}
    \caption{Recurrence time (top) and peak luminosity in units of $10^{38}~{\rm erg~s^{-1}}$ (bottom) are shown as a function of accretion rate normalized by $10^{-9}~M_{\odot}~{\rm yr}^{-1}$. The dotted line indicate the data of GS 1826$-$24 observed in 2007, assuming the typical value of $d\sqrt{\xi}_b=6.1~{\rm kpc}$ \citep{2017PASA...34...19G}, where $d$ is the distance and $\xi_b$ is the anisotropy of burst flux. 
    } 
     \label{fig:bp}
\end{figure}

\subsection{Dependence on the accretion rate and implication for GS 1826$-$24}
\label{sec:3c}

In order to see the dependence of the burst behavior on the accretion rate, in Fig.~\ref{fig:bp}, we show the recurrence time $\Delta t$ and peak luminosity $L_{\rm pk}$ as a function of accretion rate. From the top panel of this figure, one can observe that $\Delta t$ with the DU process is significantly different from that without the DU process, where the deviation becomes at most $\sim 0.5$~hr. As we have already discussed with $\dot{M}_{-9}=2.0$, if the effect of the superfluidity is taken into account, $\Delta t$ becomes slightly lower. On the other hand, the peak luminosity seems to strongly depend on the effect of the superfluidity, compared with the dependence in $\Delta t$. 

Now, we compare these results to the data of GS 1826$-$24 observed in 2007. In Fig.~\ref{fig:bp} we plot the observational data with the dotted lines. Unfortunately the mass accretion rate in GS 1826$-$24 is still uncertain, but one can distinguish the models if the mass accretion rate will be constrained somehow. For example, if $\dot{M}_{-9}\simeq 2.0$, one can say that the DU process is unfavorable for explaining the observed $\Delta t$ of GS 1826$-$24, at least within the models we considered in this study. From this comparison, we find that the DU process seems to be unfavorable for explaining the observed $\Delta t$ of GS 1826$-$24, at least within the models we considered in this study. We also compare the peak luminosities with the obsetvation of GS 1826$-$24 in 2007. For the observed peak luminosity, we take it from observed peak flux in 2007 ($f_{\rm peak}=\left(2.76\pm0.092\right)\times{\rm 10^{-8}~{\rm erg~cm^{-2}~s^{-1}}}$) and the typical observational value of $d\sqrt{\xi_b}=6.1$ kpc, where $d$ is the distance and $\xi_b$ is the burst anisotropy~\citep{2017PASA...34...19G}. In this case, the models with DU process can not explain the peak luminosity observed in GS 1826$-$24, which is the same result as the constraints from $\Delta t$. Although the theoretical results highly depends on the distance and other input parameters~(e.g., \cite{2020MNRAS.494.4576J}), we could thus extract the information of neutrino cooling processes including nucleon superfluidity through the burst observations through systematical examinations. 

Light-curve fitting with the observations of Clocked bursters is also good tool to constrain the NS core properties from X-ray burst observations but this is very hard at present (e.g., \cite{2021ApJ...923...64D}). The most uncertain parameter is the accretion rate, which is in principle difficult to be measured by observations. The uncertainties of reaction rate above all for the $\alpha p$ and $rp$ processes also hasten the difficulty of constraints from burst observations, while $\Delta t$ and $L_{{\rm pk},38}$ focused in this study are not sensitive to heavy-element nucleosynthesis. As experiments to measure the reaction rates of unstable $p-$nuclei are developed, the light-curve fitting becomes more useful tool to constrain the NS properties and accretion rate. Then, the light-curve fitting to probe the structure and temperature of NS core will be presented elsewhere.

\begin{figure}
    \centering
    \includegraphics{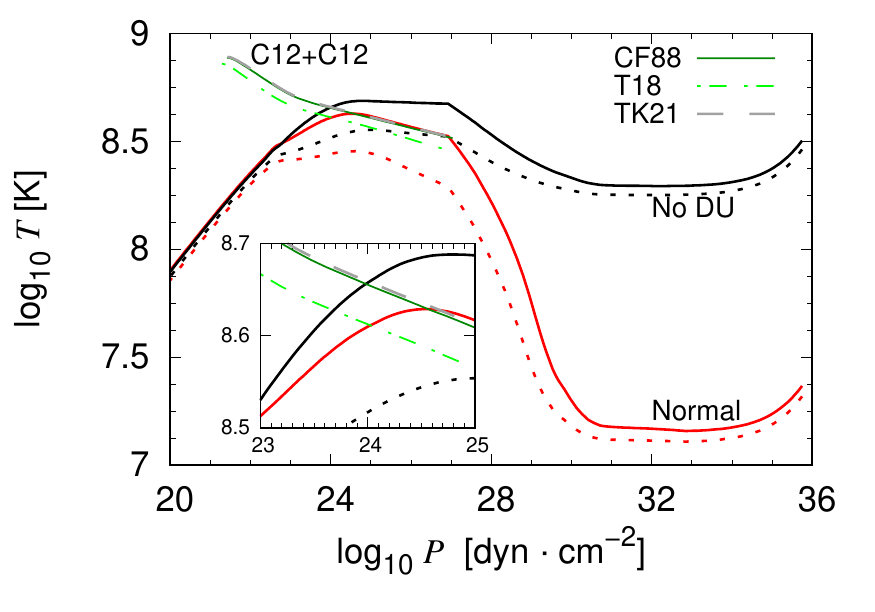}
    \caption{Temperature structure just before the H/He ignition with (red) and without (black) the DU process. Solid curves indicate $\dot{M}_{-9}=8$ while dotted curves $\dot{M}_{-9}=4$. We ignore the effect of any kind of superfluidity. We also plot three kinds of carbon ignition curves with the carbon mass fraction  $X({}^{12}{\rm C})=0.2$: CF88 \citep{1988ADNDT..40..283C}, T18 \citep{2018Natur.557..687T}, and TK21 \citep{2021PhLB..82336790T}.}
    \label{fig:c12c12}
\end{figure}

\section{The ignition and recurrence time for   superbursts}
\label{sec:c12c12}

 In this section, we investigate the effect of the DU process on the depth of carbon ignition and the recurrence time of superburst based on Eq.~(\ref{eq:eq1}). As the temperature structure inside the NS just before the carbon ignition, we choose that before the H/He ignition ($t\simeq10^6~{\rm s}$), which is the same as the temperature structure denoted with the dotted lines in Fig.~\ref{fig:istrc}. We note that the temperature profile itself is not so changed regardless of the mass fraction of carbons unless the nuclear burning works. We choose high accretion rate of $\dot{M}_{-9}=4$ and 8,  because the carbon burning should occur in hot NS layers $\approx0.6$ GK. We simply consider two extreme cases without and with the DU process, where any kind of superfluidity is neglected. Burst models are made in the same procedure as mentioned in Section~\ref{sec:setup}.

The non-resonant experimental rate is often used as the standard reaction rate of ${}^{12}\mathrm{C}+{}^{12}\mathrm{C}$~(\cite{1988ADNDT..40..283C}, hereafter CF88), while, above the threshold energy of ${}^{12}\mathrm{C}+{}^{12}\mathrm{C}$, many molecular resonances are experimentally predicted from the indirect measurement of $\alpha$-inelastic cross section~\citep{2013JPhCS.436a2009K}. The effect of resonances tends to enhance the reaction rates from theoretical models~(e.g., \cite{2009ApJ...702..660C} for resonance assumed at $E_R=1.5$ MeV). Actually, recent measurements of cross sections with use of the Trojan Horse Method showed some resonances in low energies, which increases the  reaction rate of ${}^{12}\mathrm{C}+{}^{12}\mathrm{C}$ by $\gtrsim25$ times at $T=$0.5 GK compared with CF88~(\cite{2018Natur.557..687T}, hereafter T18). The impact of the new reaction rate on astrophysical phenomena has been investigated, such as the type Ia supernovae~\citep{2019MNRAS.482L..70M} and evolution of massive stars~\citep{2021ApJ...916...79C}. Furthermore, the latest reaction rate has been recently developed based on a full microscopic nuclear model to describe low-energy resonances by handling the channel coupling and the rotation of nuclei without any adjustable parameter~(\cite{2021PhLB..82336790T}, hereafter TK21). We adopt the three kinds of reaction rates of ${}^{12}\mathrm{C}+{}^{12}\mathrm{C}$, i.e., CF88, T18, and TK21, in this study.

To obtain the ignition curve for the reaction of ${}^{12}\mathrm{C}+{}^{12}\mathrm{C}$, we employ the shell-flush model~\citep{1981ApJ...247..267F,1999A&A...342..464K,2004ApJ...603..242K} with Helmholtz EOS~\citep{2000ApJS..126..501T}. For the compositions,
we assume the carbon-iron plasma with the carbon mass fraction of $X({}^{12}{\rm C})=0.2$, which is the typical mass fraction to fit the observed superburst light curves with hydrogen accretion~\citep{2003ApJ...595.1077C,2006ApJ...646..429C} (but see also \cite{2001ApJ...559L.127C}).

\begin{table}[th]
    \centering
    \caption{Ignition pressure of carbon burning $P_{\mathrm{C}}$ and the recurrence time of superburst $\Delta t_{\rm sb}$. Note that the model with $\dot{M}_{-9}=4$ and DU process does not cross the ignition curves as seen in Fig.~\ref{fig:c12c12} irrespective of the reaction rates of ${}^{12}\mathrm{C}+{}^{12}\mathrm{C}$.}
    \begin{tabular}{ccc|cc}
    \hline\hline
    $\dot{M}_{-9}$ & Model &  Rate & $P_{\mathrm{C}}$ (${\rm dyn~cm}^{-2}$) & $\Delta t_{\rm sb}~({\rm yr})$ \\
    \hline
    8 & Normal  & CF88 & 3.49$\times10^{24}$ &  1.17$\times10^{-2}$\\
     &   & T18 & 1.06$\times10^{24}$ & 3.57$\times10^{-3}$ \\
      &   & TK21 & 4.79$\times10^{24}$ & 1.60$\times10^{-2}$ \\
    8  & No DU  & CF88 & 9.74$\times10^{23}$ &  3.26$\times10^{-3}$ \\
     &   & T18 & 4.87$\times10^{23}$ & 1.63$\times10^{-3}$ \\
      &   & TK21 & 1.04$\times10^{24}$ & 3.48$\times10^{-3}$ \\
       4 & No DU  & CF88 & 6.88$\times10^{26}$ & 4.61 \\
     &   & T18 & 1.71$\times10^{25}$ & 0.115 \\
      &   & TK21 & 9.31$\times10^{26}$ &  6.24 \\
    \hline\hline
    \end{tabular}
    \label{tab:table2}
\end{table}

In Fig.~\ref{fig:c12c12}, we show the temperature structure just before the H/He ignition, together with the carbon ignition curves. With this figure, one can determine the ignition pressure $P_{\mathrm{C}}$ as the intersection between the thermal structure and the carbon ignition curves. The obtained values of $P_{\mathrm{C}}$ and the recurrence time of superburst $\Delta t_{\rm sb}$ are shown in Table~\ref{tab:table2}, where $\Delta t_{\rm sb}$ is estimated from Eq. (\ref{eq:eq1}) together with $P_{\mathrm{C}}$. Focusing on the superburst with  $\dot{M}_{-9}=8$, we can find that the DU process makes $P_{\mathrm{C}}$ (and also $\Delta t_{\rm sb}$) larger by a factor of $\sim$2. For the case with $\dot{M}_{-9}=4$, since the DU process indirectly decreases the temperature for weaker crustal heating than in $\dot{M}_{-9}=8$, there is no cross point between the thermal structure and ignition curves with any type of reaction rate. This predicts that the carbon burning does not occur for this model (and $X({}^{12}{\rm C})=0.2$). Hence, if the DU process occurs in superbursters, the accretion rate as the heat source inside NSs is preferred to be high because of the necessity of hot regime enough to cause the carbon burning. 

Next, we focus on the uncertainties of reaction rates of ${}^{12}\mathrm{C}+{}^{12}\mathrm{C}$. As we see Table~\ref{tab:table2}, $\Delta t_{\rm sb}$ between CF88 and TK21 are not so changed even through the energy dependence of astrophysical $S$ factor is highly different between them (see Fig.~2 in \cite{2021PhLB..82336790T}). Meanwhile, T18 is higher around twice than others because the astrophysical $S$ factor of T18 is roughly higher than others by an order of magnitude. Hence, the uncertainties of reaction rates are similar with those of the DU process for $\dot{M}_{-9}=8$. 

We should note that our estimation of $\Delta t_{\rm sb}$ for $\dot{M}_{-9}=8$ is three orders of magnitude smaller than the observations of superburst such as the 4U 1820$-$30 ($\Delta t_{\rm sb}=5$--10~yr in \cite{2003ApJ...595.1077C}). This is due to the difference of ignition column density $P_{\mathrm{C}}/g_s$~\citep{2006ApJ...646..429C}. For $\dot{M}_{-9}=4$ without DU process, CF88 and TK21 mostly match with the observations of 4U 1820$-$30, but for T18, $\Delta t_{\rm sb}$ is lower by 1--2 orders of magnitude. Considering also the necessity of hot regime implying high $\dot{M}_{-9}$, although the uncertainties of crustal heating are crucial~(e.g., \cite{2021MNRAS.507.3860S,2022arXiv220207486S}), T18 seems to be unpreferred.

Many previous studies of thermal evolution of accreting NSs consider the shallow heating which is necessary for explaining some hot accreting NSs~\citep{2009ApJ...698.1020B,2015ApJ...809L..31D,2016MNRAS.456.4001W,2016MNRAS.461.4400O,2018MNRAS.477.2900O}, though their physical mechanism has been still unknown and some candidates have been investigated (e.g., \cite{2018PhRvC..98b5801F,2021PhRvD.104l3004L} for neutrino heating scenario due to charged pion decay). We turn off the shallow heating in this study, but depending on its depth and strength, such a heating could have a significant impact on the carbon ignition depth. In fact, \cite{2022arXiv220803347M} investigates this and tries to constrain the properties of shallow heating from recurrence time in some superburst observations (see Figure 4 in his paper). Probing the origin of the shallow heat source as well as the nature of dense matter in NSs are therefore valuable and left for our future work.

After submitting the first manuscript, we became aware of similar work of \cite{2022arXiv220803347M} in terms of constraints on energy sources in accreting NSs from the inferred depth of carbon ignition in X-ray superbursts. He also adopts T18 and TK21 as two of  ${}^{12}{\rm C} + {}^{12}{\rm C}$ reaction rates as we do, and his results are in good agreement with ours. He also examines the uncertainties of neutrino Urca cooling in the crust~\citep{2014Natur.505...62S}, which does not impact the carbon ignition depth of superburst, 
unless the densities where the heating and cooling sources greatly work are close each other. We do not consider the neutrino Urca cooling in the crust, but this seems not to be a problem at least in moderate accretion rates considered here because the neutrino cooling above all for the DU process is clearly dominant for the core temperature, while the crustal heating (and possibly shallow heating) for the crust temperature. 

\section{Concluding Remarks}
\label{sec:conc}

We investigate the neutrino cooling effect on the burst light curves with emphasis on the DU process and nucleon superfluidity. We show that the DU process makes the recurrence time $\Delta t$ longer and the peak luminosity $L_{\rm pk}$ higher, but this is suppressed by neutrons superfluidity if the critical temperature is relatively high. In our burst models, both of them are changed by maximally $\lesssim20\%$. In comparison with the observations of GS 1826$-$24, we present the possibility to probe the occurrence of the DU process and the strength of neutron superfluidty.

We also discuss the recurrence time of superburst estimated with the ignition pressure of carbon burning, considering the DU process and various reaction rates of ${}^{12}\mathrm{C}+{}^{12}\mathrm{C}$. We show that if the DU process occurs, it takes much time to cause the carbon burning or possibly carbon is not ignited. Hence, the observation of superburst could probe the neutrino cooling process because the carbon ignition occurs in deep layer, where the DU process affects compared with the case of H/He burning. We also show that the recent experimental reaction rate of ${}^{12}\mathrm{C}+{}^{12}\mathrm{C}$ (T18) makes $\Delta t$ lower by 1--2 orders of magnitude and becomes inconsistent with the observation in 4U 1820$-$30. For conventional reaction rate (CF88) and the latest theoretical one (TK21), we find that these difference appears to be small in ignition curves.

Specifying the neutrino cooling processes including nucleon superfluidity from burst observations is still difficult because there are many susceptible input parameters. Hence, we might need to search other possible observations. One of candidates is the gravitational wave with the gravity mode ($g$-mode), which could be trapped in radiative layers in NSs. Even if the cooling (and heating) processes inside NSs are not reflected on the observed luminosity, since they highly change the temperature structure, we could specify the cooling and heating processes through the signature of $g$-mode frequency. In fact, such studies have been done in isolated NSs cooling~\citep{2015PhRvD..92f3009K,2022PhRvD}. We leave an investigation of $g$-mode frequency in accreting NSs for a future study.

Recently, a very long duration of bursts compared with the superburst discovered in MAXI J0556$-$332 has been proposed to be triggered by heavier elements than ${}^{\rm 12}{\rm C}$~({\it hyperburst}, \cite{2022arXiv220203962P}). They suggest that the hyperburst is triggered by unstable burning of neutron rich isotopes of oxygen or neon around the high density of $\rho\approx10^{11}~{\rm g~cm^{-3}}$. Thus, the heating and cooling processes inside NSs including the DU process should effectively work and finally change the ignition conditions of triggers for the hyperburst. As with the superburst discussed in this paper, the observations of future very long duration bursts would be useful for specifying heating and cooling processes, were the scenario of hyperburst correct.

\small
\section*{Acknowledgments}
The authors thank Y.~Taniguchi and M.~Kimura for discussion on the ${}^{12}{\rm C} + {}^{12}{\rm C}$ reaction rate and for providing the data. This project was financially supported by JSPS KAKENHI (19H00693, 19KK0354, 20H05648, 20H04753, 21H01087, 21H01088) and a RIKEN pioneering project ``Evolution of Matter in the Universe (r-EMU)''. N.N. was supported by the Incentive Research Project at RIKEN. H.L. was supported financially by the National Natural Science Foundation of China under No. 11803026 and Xinjiang Natural Science Foundation under No. 2020D01C063. T.N. acknowledges the support from the Discretionary Budget of the President of Kurume Institute of Technology. Parts of the computations in this study were carried out on computer facilities at CfCA in National Astronomical Observatory of Japan.

\appendix
\section{On the cooling effect on burst parameter $\alpha$}
\begin{figure}[h]
    \centering
    \begin{minipage}{0.49\linewidth}
     \includegraphics{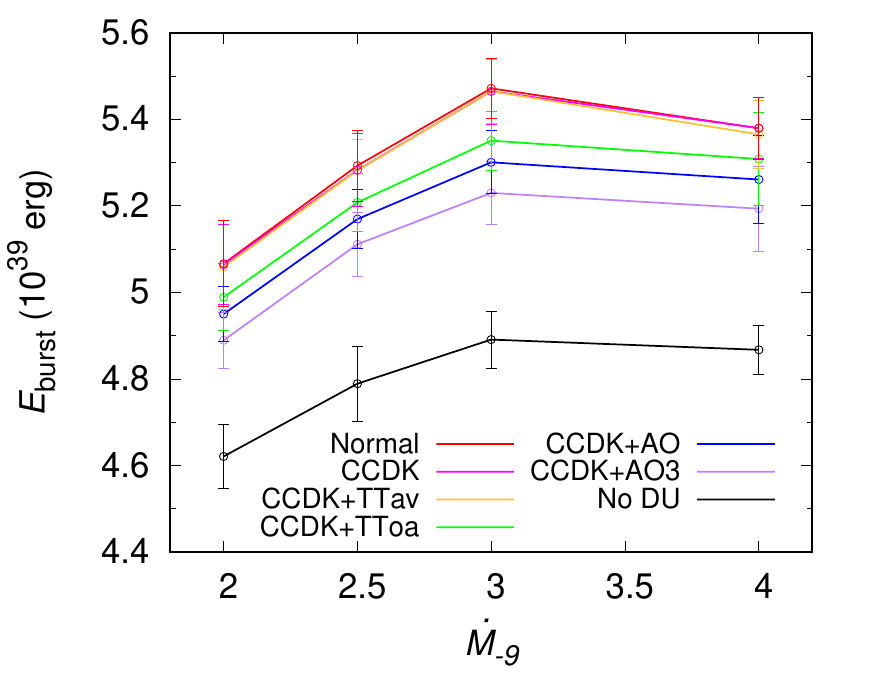}
    \end{minipage}
    \begin{minipage}{0.49\linewidth}
    \includegraphics{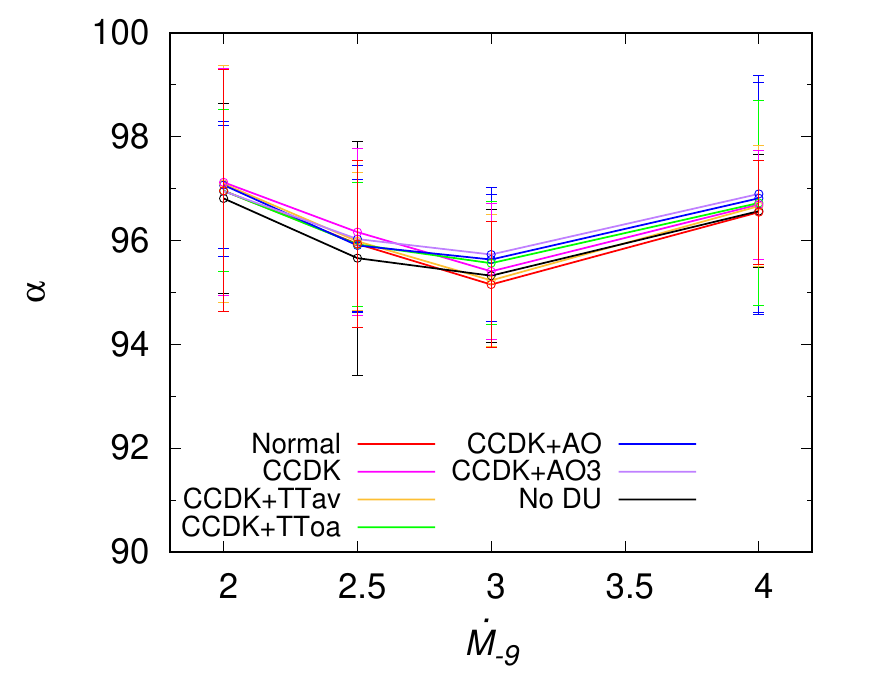}
    \end{minipage}
    \caption{Same as Fig.~\ref{fig:bp}, but for total burst energy (left) and burst parameter $\alpha$ (right).}
    \label{fig:alpha}
\end{figure}

As with the recurrence time $\Delta t$ and peak luminosity $L_{\rm peak}$, the burst parameter $\alpha$, which is the
ratio of the accretion energy in a cycle of burst to the total burst energy $E_{\rm burst}$, is also powerful for constraining burst models. It can be calculated by 
\begin{eqnarray}
    \alpha = \frac{z_g}{1+z_g}\left(10^{-9}~M_{\odot}~{\rm yr}^{-1}\right)\dot{M}_{-9}c^2\frac{\Delta t}{E_{\rm burst}},~\label{eq:eq2}
\end{eqnarray}
where $z_g$ is the gravitational redshift~\citep{2017PASA...34...19G}.  In Fig.~\ref{fig:alpha} we show the neutrino cooling effect on $E_{\rm burst}$ and $\alpha$ as a function of the accretion rate. As we see, $\alpha$ value is almost the same regardless of neutrino cooling models. This is because the tendencies of cooling effect on $E_{\rm burst}$ and $\Delta t$ are almost the same. Thus, we cannot extract the information of neutrino cooling from $\alpha$. This feature is consistent with our previous work~\citep{2021ApJ...923...64D}, in which $\alpha$ has a positive correlation with surface gravity with high accuracy and therefore is more powerful for constraining NS structure than other output values such as $\Delta t$ and $L_{\rm pk}$.

\bibliography{ref}{}
\bibliographystyle{aasjournal}



\end{document}